\documentclass{sig-alternate}

\begin{document}
%
\conferenceinfo{DISIO 2011}{March 21, Barcelona, Spain.}
\CopyrightYear{2011} 
\crdata{}  

\title{Rendering of 3D Dynamic Virtual Environments
}
%
%
%
%
%

\numberofauthors{4} 
%
\author{
%
%
\alignauthor
Salvatore A. Catanese\\
	\affaddr{Dept. of Physics\\ Informatics Section}\\
	\affaddr{University of Messina, Italy}\\
	\email{salvocatanese@gmail.com}
\alignauthor
Emilio Ferrara\\
  \affaddr{Dept. of Mathematics}\\
  \affaddr{University of Messina, Italy}\\
  \email{emilio.ferrara@unime.it}
\alignauthor
Giacomo Fiumara\\
  \affaddr{Dept. of Physics\\ Informatics Section}\\
  \affaddr{University of Messina, Italy}\\
  \email{ giacomo.fiumara@unime.it}
\and \alignauthor 
Francesco Pagano\\
  \affaddr{Dept. of Information Technology}\\ 
  \affaddr{University of Milan, Italy}
  \email{francesco.pagano@unimi.it}}

\date{}

\hyphenation{video-game video-games CAE-LUM-COM-PONENT-SKY-DO-ME CAE-LUM-COM-PONENT-POINT-STAR-FIELD CAE-LUM-COM-PONENT-CLOUDS CAE-LUM-COMP-ONENT-MOON CAELUM-COMPONENT-SUN Game-Character-Controller}

\maketitle
\begin{abstract}

%

In this paper we present a framework for the rendering of dynamic 3D virtual environments which can be integrated in the development of videogames.
It includes methods to manage sounds and particle effects, paged static geometries, the support of a physics engine and various input systems.
It has been designed with a modular structure to allow future expansions.

We exploited some open-source state-of-the-art components such as \textit{OGRE}, \textit{PhysX}, \textit{ParticleUniverse}, etc.; all of them have been properly integrated to obtain peculiar physical and environmental effects.
The stand-alone version of the application is fully compatible with \textit{Direct3D} and \textit{OpenGL} APIs and adopts \textit{OpenAL} APIs to manage audio cards.

Concluding, we devised a showcase demo which reproduces a dynamic 3D environment, including some particular effects: the alternation of day and night influencing the lighting of the scene, the rendering of terrain, water and vegetation, the reproduction of sounds and atmospheric agents.

\end{abstract}




\section{Introduction}

The design of virtual environments is a complex and expensive process within the development of videogames.
The improvements in programming video cards greatly increases the possibility of creating extremely involving interactive virtual environments and worlds with enhanced exploratory choices in immersive playing experiences.

Besides the generation of virtual object models such as buildings and landscapes and a real-time camera control for navigation in virtual environments, the development should provide the possibility to manage physical, environmental and collision effects.
These elements determined a substantial increase of development costs in order to face the large amount of extra technical and artistic features.
The result is that, in some cases, production costs are comparable to those of some film productions.

Although some virtual environment development toolkits are available, some of them provide only a subset of those instruments which are necessary to completely develop virtual worlds.
Some features particularly difficult to be simulated, such as wind, fire, smoke and water often need a further programming phase, thus still increasing development costs.

In the last years there has been an increase in the number of middlewares and frameworks (see further) which try to face the problem of fulfillment and optimization of 3D dynamic virtual environments in order to solve the technical requirements of complex videogames.
The framework we developed can be set in this category.

The virtual environment we describe in this paper is build on top of a previously developed middleware framework \cite{catanese2011middleware} with features to manage input devices,  sound and music integration, networking support and physics effects implementation. 
This work extends the features of this platform improving the graphical quality, the degree of realism and involvement of virtual scenes, when non-commercial engines and libraries are employed.

\section{Related work}

A lot of research has been conducted, both by academia and enterprises, in supporting videogames development in reproducing 3D virtual environments.
In the last few years, the most of the efforts have been addressed in reproducing realistic virtual environments which include those phenomena usually characterizing real environments, such as the physics, environmental effects, photo-realistic graphics, and so on.
Our work focuses on some of these aspects and most of them have been already covered in specific related work.

Because developing virtual environments fuses aspects of software engineering, architecture, artificial intelligence, 3D graphics, art and sound effects, frameworks and platforms supporting these steps have been developed in the last years.
For example, Trenholme and Smith in \cite{trenholme2008computer} present an overview of several currently available commercial game engines (\textsl{id Tech 3-4}, \textsl{CryENGINE}, \textsl{Source} and \textsl{Unreal Engine 2}), which are suitable for prototyping virtual environments.
Similarly, Watson et al. \cite{watson2008procedural} review the procedural modeling, examining the \textsl{CityEngine} game engine they developed, and studying the use of procedural urban modeling in Electronic Arts' \textsl{Need for Speed} games, for representing virtual urban environments exploiting their middleware.
On the other hand, in the academic field, a simple game engine (\textsl{SAGE}), developed in a game programming class at the University of North Texas, is described by  Parberry et al. in \cite{parberry2007sage}. 
They show a sequence of demos implementing different functionalities; each demo extends its predecessor in a process called incremental development.
\textsl{SAGE} generates 3D virtual environments, that users can explore in real time, containing interactive objects. 
It includes a graphics renderer, using pixel shaders and HLSL, objects, terrain and some method for level-of-detail to increase rendering speed, input management and collision detection.
\cite{parberry2007sage} shares similarities with our work in these aspects.

In the area of environmental effects simulation, several methods for reproducing weather phenomena, like particle-based rain techniques, are presented in \cite{tatarchuk2006artist}. 
The authors describe in details those methods to render, in a real-time system, very complex atmospheric physical phenomena such as strong rainfall, falling raindrops dripping off objects' surfaces, various reflections in surface materials and puddles, water ripples and puddles on the streets and so on.

Realistic animations of water, smoke, explosions, and related phenomena are reproduced via fluids simulation.
Wicke et al. \cite{wicke2009modular} face the problem of high-resolution fluid motion in real-time videogame applications, describing some techniques on a scale previously unattainable in computer graphics. 
The central idea is to cover the simulation domain with a small set of simulation primitives, called tiles. 
Each tile consists of a velocity basis representing the possible flow within its sub-domain. 
The boundaries between sub-domains correspond to tile faces.

Finally, in the area of realistic graphics, an exterminated amount of work has been presented in the last years.
An interesting interactive fractal landscape synthesizer on programmable graphics hardware, which exploits the intrinsic strengths of GPUs to generate and render high-quality, high-resolution, textured and shaded terrains has been presented by Schneider et al. in \cite{schneider2006real}. 
The synthesis step is directly integrated into the rendering procedure and requires neither any polygonal representation nor a pre-processing stage. 
The algorithms are combined into a visual interface to allow the intuitive design of highly detailed terrain models.

An overview of \textsl{Halo 3}'s unique lighting and material system and its main components is treated in \cite{chen2008lighting}. 
Halo includes key innovations in the following areas: spherical harmonics lightmap generation, compression and rendering; rendering complex materials under area light sources; HDR rendering and post-processing. Some of the effects presented in that work have been adopted also here (e.g. the HDR rendering).

\section{The Framework}

The application we developed includes some of the most important open-source solutions used for the rendering of 3D virtual environments.
We employed: \textsl{OGRE} (Objected-Oriented Graphics Rendering Engine) \cite{ogre3d}; \textsl{Caelum} for photo-realistic creation of the sky, clouds and atmospheric agents; \textsl{Hydrax} for the rendering of scenes including water and for the reproduction of its effects such as depth, foam, sunbeams through water surface, etc.

\textsl{OGRE} was chosen among other open-source engines for various reasons.
Some graphical engines while showing a large list of features, can be hardly merged to create a usable tool.
Others are supplied with enchanting demos, but are scarcely helpful when a broader project is to be created.
Finally, others are too much specific for a given type of videogame.
What makes \textsl{OGRE} different from other graphical engines is the wide support community, the documentation and its being open-source.

We have used \textsl{Paged Geometry} for the rendering of large amounts of little meshes needed to cover the surfaces of the environment: the library is specifically designed for the rendering of forests and outdoor scenes where mushrooms, rocks and blades of grass must be rendered in a performing way.
The \textsl{PhysX} engine has been chosen to dynamically simulate objects, collisions with the terrain and the static geometry (paginated or not) of the scene.
\textsl{PhysX} is also suited to be integrated with \textsl{OGRE} via the \textsl{NxOGRE} wrapper class.
We have also used the \textsl{OGREOggSound} library in order to manage sounds, together with \textsl{Particle Universe} to create visually stunning particle systems.

We also introduced two techniques, namely ``texture splatting'' and ``parallax mapping'', which improve the quality and the realism of the reproduced environment based  on programming shaders.
They will be described in detail in next sections.
Our application extends a middleware framework presented elsewhere \cite{catanese2011middleware} which provides a starting point to ease the process of game development, discussed in the next section.

\subsection{Basic Middleware Framework}

The middleware framework at the core of the application consists of a series of tools which greatly eases the development of videogames based on 3D environments.
It allows programmers and artists to focus on game dynamics and gameplay neglecting the technical aspects of development.

The system provides some features which ease the phases of development and in particular to: i) load a 3D scene created using graphical modeling softwares (e.g. \textsl{Blender}); ii) specify the physical qualities of an object (mass and physical model) inside a scene; iii) export the 3D environment towards third part applications; iv) accomplish the rendering of the dynamic environment; v) manage the motion of the objects thus accomplishing a physical simulation, and vi) save the videogame progress for further restore.

Modularity is the main characteristic of the middleware framework. 
It permits extensions and customizations for various uses, as in the case of the application we illustrate in this paper.
Each of the packages it is composed of, which can be replaced upon choice, deals with a specific functionality.
Therefore it is possible to customize the framework without having to re-implement all the characteristics if the interface of the replaced package is maintained.

The main packages are those related to the management of the cycle of rendering (GameSystem), the I/O (GameIO), the interface with audio libraries \textsl{OpenAL} (GameAudio) and the control system of the character based on \textsl{PhysX} (GameCharacterController). 
They also include the methods to load and save the scenes (GameSceneLoader).

\subsection{Framework Extended Modules}

The new modules we developed allow to accomplish the 3D rendering of a dynamic environment which includes the following features: day-night cycle simulation, to manage the lighting of the whole scene; realistic rendering of the water, with refraction and reflection effects on the objects in the scene; optimized rendering of static paged geometry; optimized and simultaneous rendering of the trees, grass and other static elements; free camera with variable velocity to explore the environment.

Moreover, the system supports: the terrain rendering and generation from ``heightmaps''; collision detection via \textsl{PhysX}; terrain texturing using the ``texture splatting'' technique; ``parallax mapping'' on road texture for good depth effect; multiple sounds playing; multiple particle system effects.

We introduced the possibility of managing random weather conditions, with dynamic wind variables (speed, direction); rain effects with different intensity, speed, direction (wind correlated); thunderstorm with lightning effects; sounds of thunders calculated according to the distance between the lightning and the observer, with real attenuation/distance retard to hear them; water surface update in base of weather conditions (calm or rough sea).

The stand-alone framework, written in C++, may be further extended to include third part libraries; extensions should be interfaced with existing modules, and eventually refactoring the replaced libraries.
It is compatible with \textsl{OpenAL}, \textsl{Direct3D} e \textsl{OpenGL} APIs, and supports a large number of video cards with 2.x pixel shader functionality.

\section{Dynamic Virtual Environments}
The process of designing of virtual environments is build on top of a previously developed middleware framework \cite{catanese2011middleware}; this platform represents a solid starting point for the development of 3D games, including several features (e.g. input devices management, sound and music integration, networking support, physics effects implementation, etc.).

In this work we extend the features of this platform, in particular improving graphics quality aspects (introducing new rendering techniques) and the realism and the degree of credibility and involvement the designed virtual environments could ensure (introducing realistic terrain generation techniques, weather management, particle systems, fluid dynamics, load balancing of renderable elements, etc.).

\subsection{Design of the Demo Environment}
\begin{figure}
	\includegraphics[width=\columnwidth]{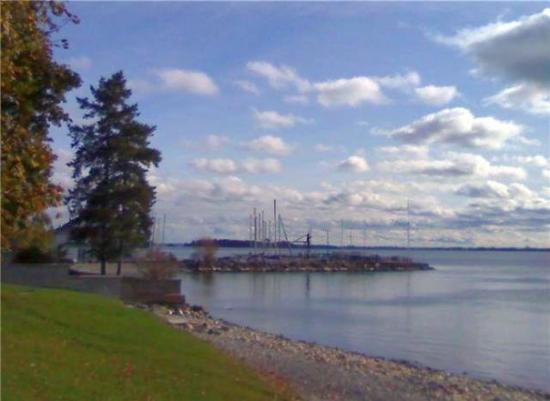}%
	\caption{Port Royal Bay (Jamaica) nowadays}%
	\label{port-royal-bay}%
\end{figure}

We also designed a demonstrating environment, which integrates the previously listed features and is used as show-case for illustrating them.
This virtual area reproduces the Port Royal Bay (Jamaica) (Figure \ref{port-royal-bay}); the place represents a realistic setting for a possible fantasy/adventure show-case demonstration; this because, during the 17th century, it was one of the main scenarios of piracy in the Caribbean Sea. 

It was necessary to reproduce a realistic environment also with respect to the temporal period.
It includes two main areas: i) the city center (including buildings, the citadel, etc.); ii) the bay (shores, plains, forests, etc.).

We adopted two different scene managers to represent the areas: for the outdoor environment we use the \textsl{TerrainSceneManager}, while for indoor places we adopt the \textsl{BSPSceneManager} (both of them are optimized for their different purposes).
A small citadel has been designed via \textsl{Blender}. 
The scene has been exported through an \textit{ad-hoc} plug-in which exploits the \textsl{DotScene} extended DTD we previously presented \cite{catanese2011middleware}.
The outdoor environment has been created using a terrain generation algorithm (details follow) starting from a heightmap taken from NASA altimetry satellites (Figure \ref{maps}, top left).

\subsection{The Terrain}
The terrain generation problem is crucial for the creation of a realistic virtual environment.
Several commercial products already exist (e.g. \textsl{Vue 9 Infinite Terrains} \footnote{http://www.e-onsoftware.com/}), which produce realistic outdoor areas via different techniques.
Our purpose is to integrate in our framework a feature for obtaining similar results.

\subsubsection{Terrain Generation via Maps}
The \textsl{TerrainSceneManager} provided with \textsl{OGRE} supports the generation of outdoor terrains via maps.
In particular, a ``heightmap'' is used to define the topology of the terrain; this map is a grayscale image in which each pixel describes the altimetry of a corresponding terrain point (Figure \ref{maps}, top left).
After the terrain topology generation, textures are applied to it (Figure \ref{textures}); then a base color is passed over the surfaces in order to diversify different areas, adopting a ``color map'' (Figure \ref{maps}, bottom right); finally, a ``detail map'' (Figure \ref{maps}, bottom left) is applied, which includes RGB and alpha channels (used for texture splatting and parallax mapping via shaders, details follow).

\begin{figure}
	\includegraphics[width=\columnwidth]{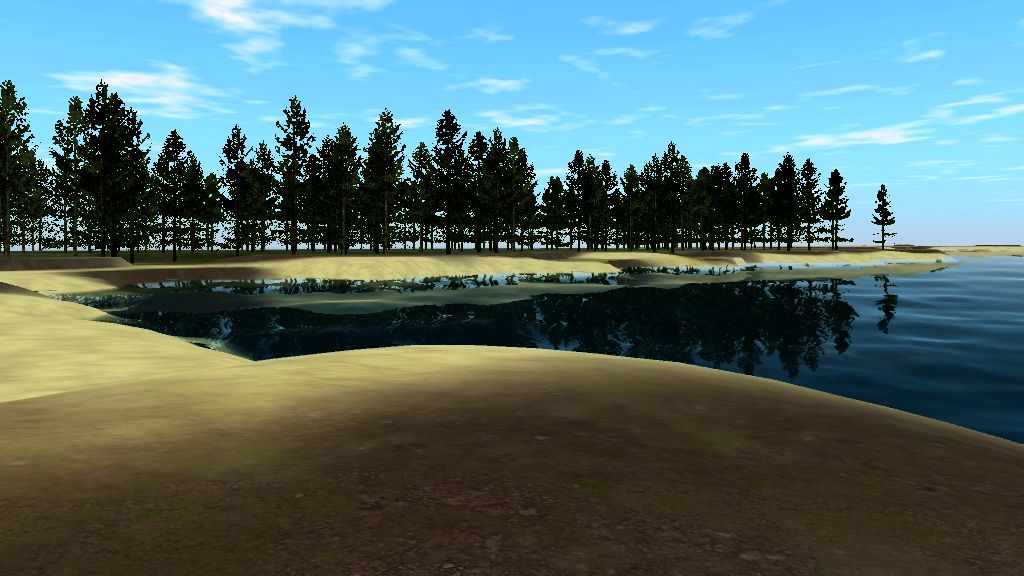}%
	\caption{3D virtual reconstruction of the bay}%
	\label{hdr8_preview}%
\end{figure}

Figure \ref{hdr8_preview}, so as several following screenshots, shows the results for our show-case example; comparing these pictures with real photos (e.g. Figure \ref{port-royal-bay}) we observe the excellent degree of realism of the reproduced environment.

\subsubsection{Maps and Texturing}
\begin{figure}
	\centering
	\includegraphics[width=150px]{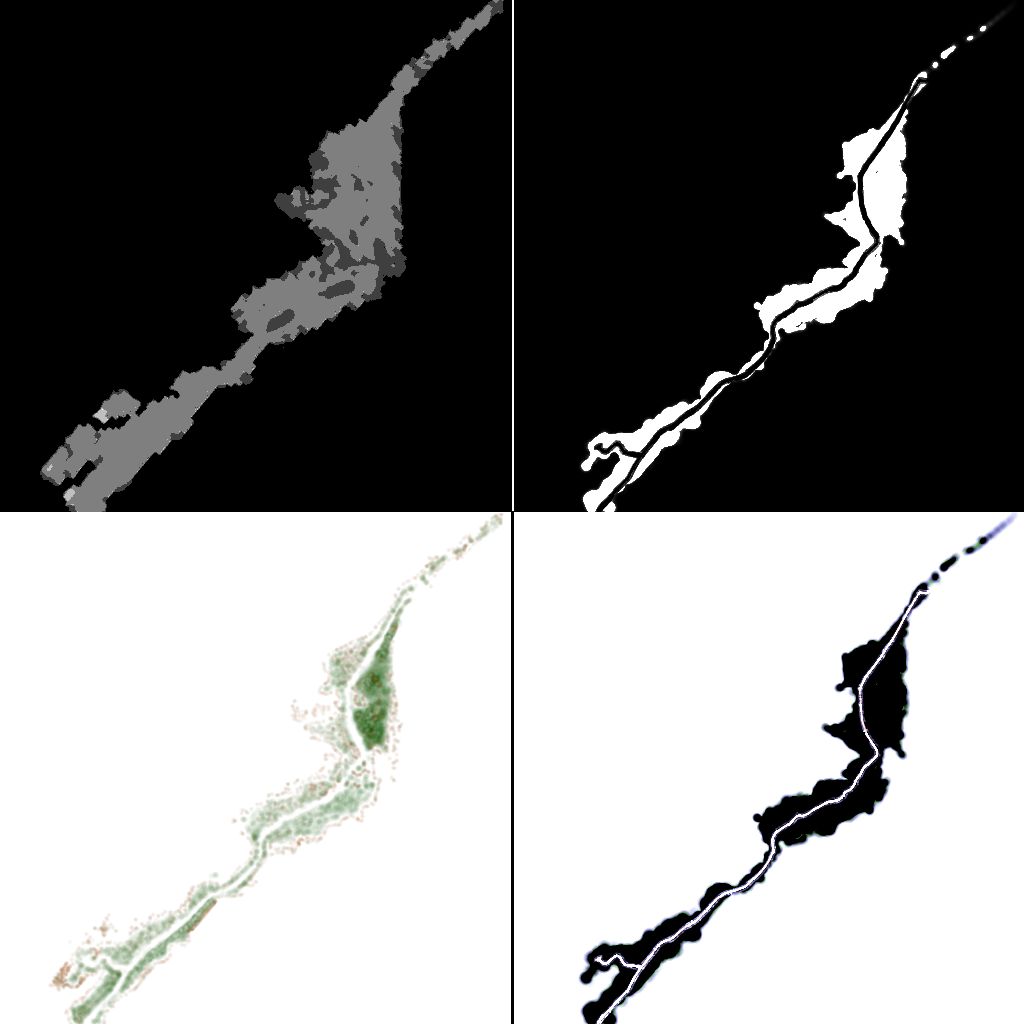}%
	\caption{Clockwise from top left: ``heightmap'', ``density map'', ``coverage map'', ``color map''}
	\label{maps}%
\end{figure}

For our show-case demonstrative environment, four different maps have been generated for creating the outdoor area (Figure \ref{maps}): i) the ``heightmap'' (previously discussed); ii) a ``density map'', which is adopted for adding entities like trees, grass, etc.; iii) a ``coverage map'' (containing information about the distribution of textures in the four channels, RGB and alpha) and iv) the ``color map'' (the base color passed on surfaces), used for texture splatting and parallax mapping processes.

\begin{figure}
	\includegraphics[width=\columnwidth]{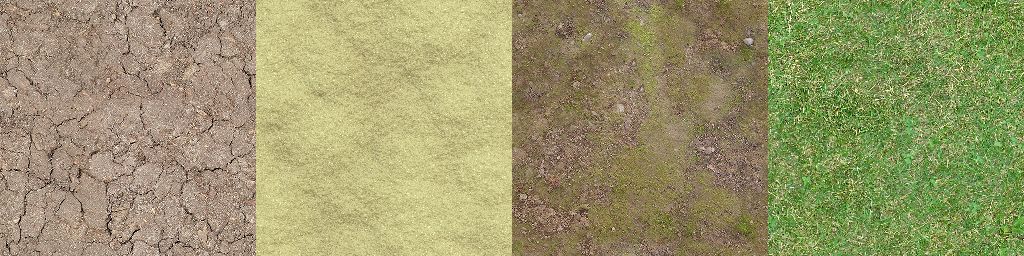}%
	\caption{Textures for the terrain}%
	\label{textures}%
\end{figure}

Four textures have been applied to the terrain (Figure \ref{textures}): i) pathway, ii) sand, iii) sand and grass and iv) short grass.

\subsection{Rendering Techniques}
The framework \cite{catanese2011middleware} relies on \textsl{OGRE} \cite{ogre3d} for rendering the virtual environment.
This open-source rendering engine provides techniques and methods for representing 3D virtual scenes, exploiting the power of GPUs (Graphics Processing Units), supporting both the \textsl{Direct3D} and the \textsl{OpenGL} pipelines.

Although several state-of-the-art algorithms have been already implemented within the rendering engine, we introduced two techniques, namely \emph{texture splatting} and \emph{parallax mapping}, which improve the quality and realism of the reproduced environment.
These techniques are implemented via shaders.

\subsubsection{Texture Splatting}
The process of \emph{texture splatting} was originally described  by Bloom \cite{bloom2000texture} in 2000. 
The purpose of this technique is to merge several textures on a unique surface via alphamaps. 
An alphamap is a grayscale image which is included in the alpha channel of a texture.

Each pixel of the alphamap determines the degree of opacity or translucency of the pixel in the same position of the texture.
In this technique the alphamaps are exploited to establish how much opaque or translucent has to be each pixel of the texture in the corresponding position within the alphamap. 
This is easily obtained by multiplying the alpha channel by the RGB channels. 
Alpha lies in the interval [0,1] (0 corresponds to completely transparent, 1 to completely opaque).

\begin{figure}
	\includegraphics[width=\columnwidth]{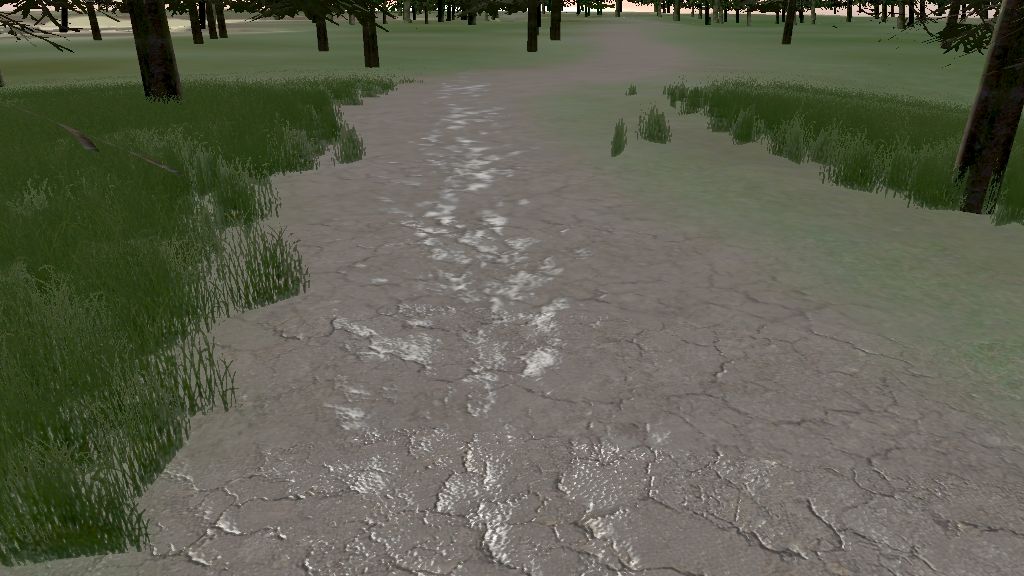}%
	\caption{Texture splatting and parallax mapping}%
	\label{hdr2_preview}%
\end{figure}

We extended the graphical features of the rendering engine implementing the texture splatting algorithm; it was adopted (together with \emph{parallax mapping}), for example, to better reproduce the effect of pathways through the forest (Figure \ref{hdr2_preview}).

\subsubsection{Parallax Mapping}
This rendering technique has been introduced by Kaneko \cite{kaneko2001detailed} in 2001, and represents an improvement of standard bump mapping algorithms; because of the computational overhead introduced, this technique has been widely exploited only in the last few years thanks to the adoption of GPUs programming.
The parallax mapping is implemented using normal maps for displacing the position of \emph{texture coordinates} on the rendered polygon, considering the value of the normal map in each point of the texture, as a function of the view angle in tangent space.

As shown in Figure \ref{hdr2_preview}, due to the parallax effect, this technique improves the realism of texturing flat surfaces, giving the optic illusion of depth.
The rendering engine has been extended for supporting the parallax mapping with multiple iterations of the process.

\subsection{The Skydome}
In order to create an illusion of being projected into an environment bigger than it really is, our framework includes the possibility of embedding the whole world inside a box, namely the skydome. 
The skydome represents a technique used for creating backgrounds.
This way, distant and unreachable objects such as the sky, boundary mountains, etc. are projected onto this container, thus obtaining the optic illusion of three-dimensional surroundings.

The skydome we adopted is a hemisphere (instead of a cube, as usual), in order to improve the degree of realism.
The skydome could be fixed, but a realistic virtual environment should include the possibility of reproducing dynamic skydomes.
Our system introduces a novel approach to simulate the day-night cycle, also including a dynamic representation of clouds, sun/moon lighting and realistic positioning, astronomically correct starfield, and much more.

\subsubsection{The Caelum Library}
\textsl{Caelum} is an open-source library developed as a plug-in for \textsl{OGRE} which aims to support a photo-realistic representation of the skydome.
It introduces the possibility of managing objects such as the sky (e.g. dynamic color modification), clouds, sun and moon, the starfield, etc.
Through the integration of this library within the framework we obtained a more attractive representation of the sky.

Several modifications to the core of this library have been introduced in order to: i) integrate \textsl{Caelum} with the \textsl{Weather} library we developed (e.g. to produce more realistic atmospheric events); ii) exploit the particle system (e.g. for generating clouds); iii) communicate with the \textsl{Hydrax} library for interchanging information about the position of the sun/moon (e.g. for generating effects of reflection of lighting on the water).

\subsubsection{CaelumManager}
\textsl{CaelumManager} is the class we developed to integrate the functionalities of the \textsl{Caelum} library, which works as a wrapper. 
This class provides methods to access objects and properties required to implement the elements we previously described.
It integrates methods for communicating with the \textsl{WeatherManager}, for managing the clouds, and with the \textsl{HydraxManager}, for reproducing the effect of interaction among the sun/moon, skydome and starfield with the water.
Some additional minor modifications include interfacing this library with the scene manager and with the camera, for setting/getting the visibility of the sun or the moon with respect to the position of the character and the camera in the environment.

\subsubsection{The Day-Night Cycle}
After instantiating, the \textsl{CaelumManager} loads a set of parameter for initializing and managing related objects.
Components managed by the library, after our improvements, are the followings: i) CAELUM\_COMPONENT\_SKY\_DOME; ii) CAELUM\_COMPONENT\_SUN; iii) CAELUM\_COMPO-NENT\_CLOUDS; iv) CAELUM\_COMPONENT\_MOON and, finally, v) CAELUM\_COMPONENT\_POINT\_STARFIELD.

Because of our custom implementation of \textsl{WeatherManager}, we do not manage the weather component included by default in the \textsl{Caelum} library.
Several parameters are initialized, for example the \emph{time scale}, which is a multiplier for representing the time passing in the game.

\begin{figure}
	\includegraphics[width=\columnwidth]{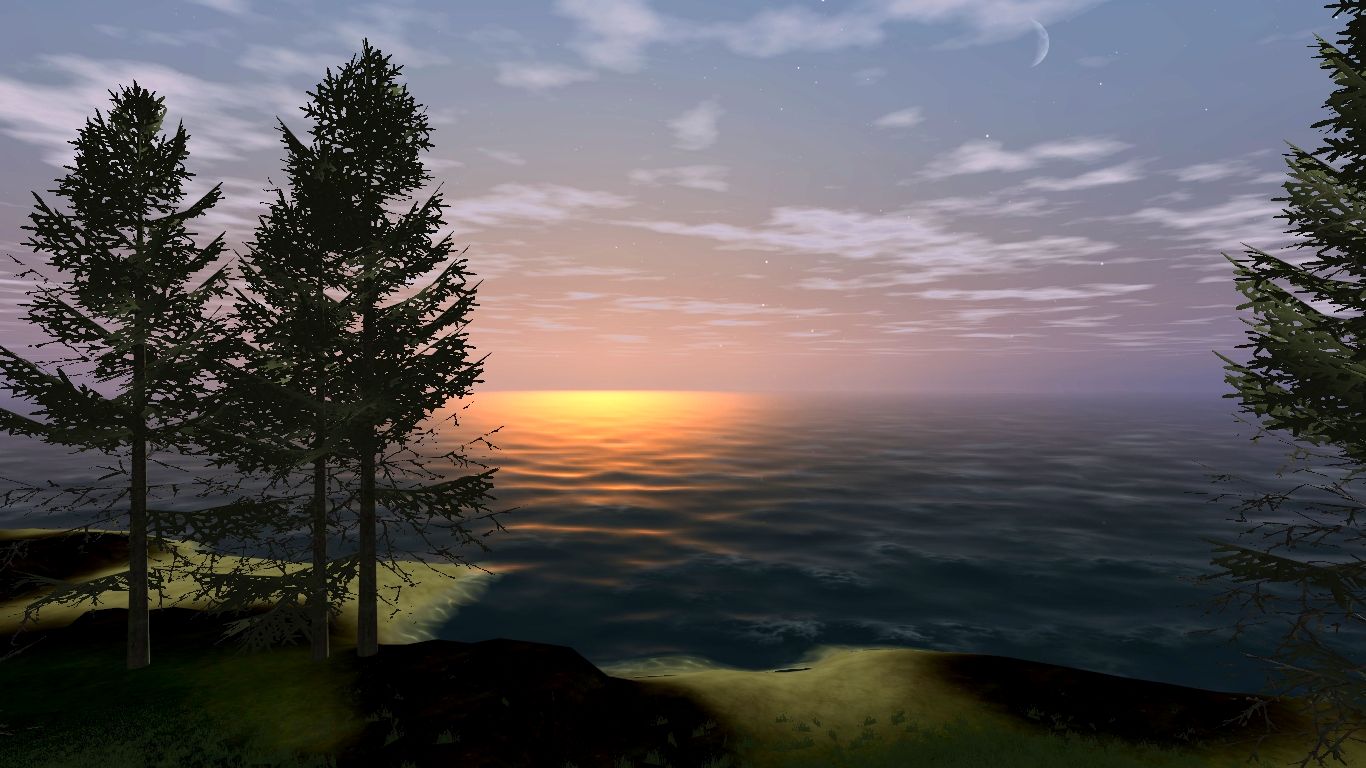}%
	\caption{The day-night cycle simulation}%
	\label{hdr9_preview}%
\end{figure}

In our demo show-case this value is set to 8, so as representing 1 minute in real as 8 minutes in game.
We introduce also the possibility of setting latitude and longitude of the environment positioning, and the Gregorian calendar date of the simulation.
In order to simulate the natural lighting we have also set a range for the ambient light, lying in the interval from the darkest blue (for the night) to the lightest yellow (for the day).
Figure \ref{hdr9_preview} depicts a dusk panorama, while several other figures within this section illustrate different condition of lighting.

\subsection{The Particle System}
A particle system is helpful to simulate several fuzzy phenomena (e.g. fire, water, fog, rain, clouds, etc.) otherwise not easily and realistically reproducible with standard rendering techniques.
There are several tools to render particle systems, usually adopting scripts to initialize, start and stop certain particle effects, able of contemporary manage different objects on the scene, hierarchies, etc.

Simulated particle systems should also be subjected to physics in order to reproduce a more realistic virtual world.
Thus, in our project we implemented a particle system manager exploiting an existing library, namely \textsl{ParticleUniverse}.
We integrated, through some modifications, its functionalities with those provided by other components of the framework.
In our demo show-case this system is mainly used to create clouds, rain, hail and snow (see Figures \ref{hdr10_preview} and \ref{hdr7_preview}).

\subsubsection{The ParticleUniverse Library}
\textsl{ParticleUniverse} is a library which can be incorporated inside the \textsl{OGRE} rendering engine to support particle systems.
This plug-in consists of an efficient run-time dynamic-link library, developed in C++, which enable the rendering engine to simulate particle systems running specific scripts.

We exploited this library for generating realistic weather effects (e.g. rain, snow, clouds, etc.).
Moreover, we introduced some modifications in order to integrate this library with \textsl{Caelum}, \textsl{Hydrax} and \textsl{PhysX}, for representing dynamics among particle systems and the physics.

\subsubsection{ParticleListManager}
\textsl{ParticleListManager} is the simple wrapper class we developed to integrate the library in our framework.
It is an efficient manager of the list of particle systems instantiated at a given time in the virtual world.
This class provides the methods to manage existing particle systems and to dynamically create new ones.

Moreover, it includes the interface to \textsl{NxOGRE} to manage the physics of interactions among particle systems, or between a particle system and other physical objects on the scene.

\subsection{The Weather}
Commercial middleware tools like \textsl{Simul Weather} \footnote{http://www.simul.co.uk/weather/} provide realistic simulation and management of the weather condition within the reproduced virtual worlds.
Usually some commonly supported features are the possibility of simulating clouds and related atmospheric effects (e.g. rain, snow, lightning, etc.).

During our development we noticed a lack of open-source products to this purpose.
Thus, we decided to develop a new library from scratch, to be included as a plug-in in \textsl{OGRE}, but virtually independent from the platform in which it should run.
This library should be integrated with the particle system, because atmospheric effects are better reproduced this way; it should also be connected to the physics engine, in order to reproduce physics effects; finally an interface to the scene manager is required to locate effects on the scene.

\subsubsection{WeatherManager}
Our \textsl{WeatherManager} class provides methods to manage the simulation of atmospheric effects on the scene.
It extends the \textsl{OGRE:FrameListener} class overloading several important methods; it includes a list of particle systems (via \textsl{ParticleUniverse}), provides several attributes for personalizing the weather simulation and, finally it instantiates a couple of objects; the three most important objects are: i) Clouds; ii) Lightning and, iii) Rain.
Its most important methods include the fader between different atmospheric condition, the wind simulator and the random precipitation generator.
Finally, two methods for saving and loading the state of the system have been included.

\subsubsection{Clouds and ManageClouds}
\begin{figure}
	\includegraphics[width=\columnwidth]{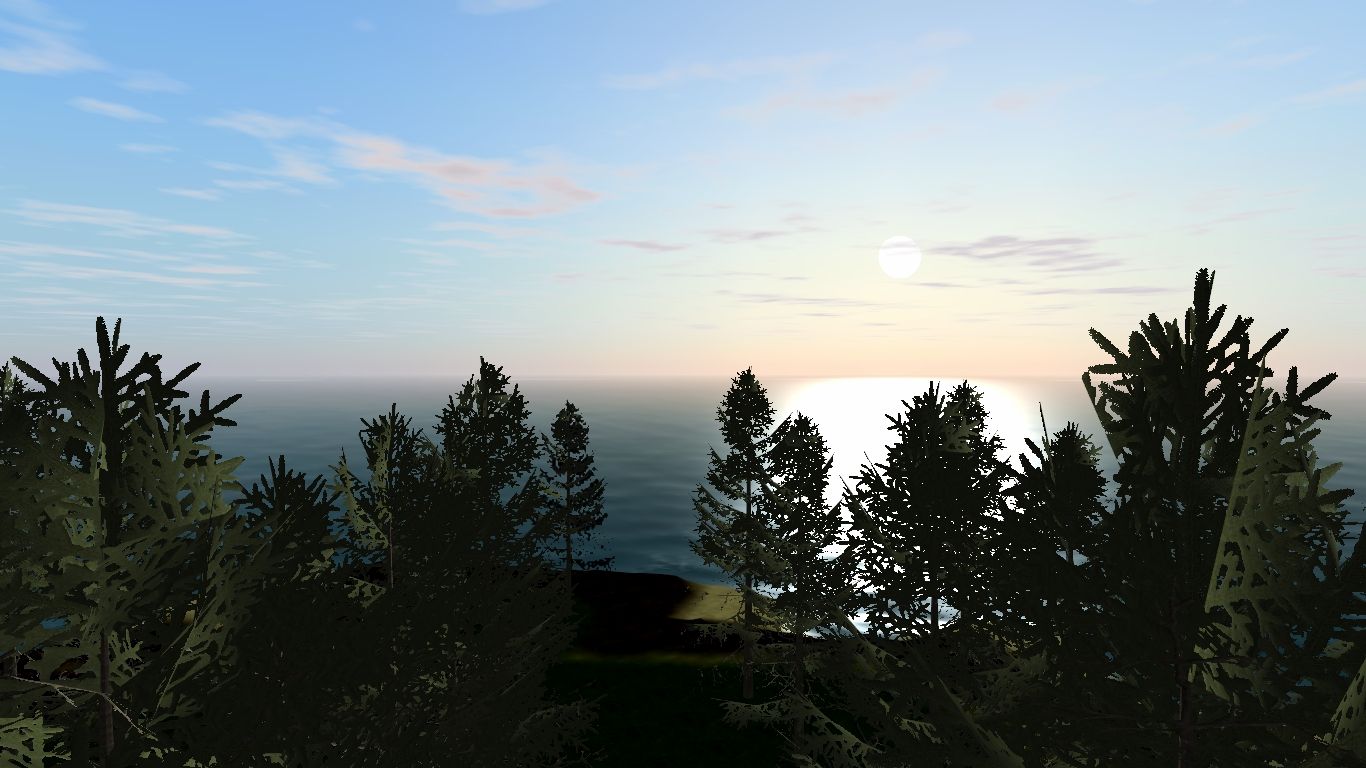}%
	\caption{Clouds and the skydome}%
	\label{hdr10_preview}%
\end{figure}

The \textsl{Clouds} class implements the cloud systems. 
Clouds can be distributed over an arbitrary number of different layers.
This class provides generative methods for instantiating a new cloud system, setting several attributes (e.g. the coverage of the clouds, the starting speed of each layer, etc.) and the methods for dynamically managing the system in real-time.
The realistic effect of transition from an atmospheric condition to another is obtained using some \textsl{fading} methods.
Clouds strictly depend from the simulation of the weather condition (e.g. the wind) reflecting these changes, and additionally from some dynamically random calculated parameters which reproduce humidity, chance of raining, etc.
In Figure \ref{hdr10_preview}, as in several other snapshots in this section, we show examples of the cloud system.

\subsubsection{Lightning and ManageLightning}
The \textsl{Lightning} class implements lightning and thunders.
Each instance represents a single lightning event, with the related thunder.
The lightning is managed as a particle system, while the thunder is just the associated sound event.

The sound of thunders is reproduced with a time delay calculated with respect to the distance of the lightning from the camera.

\subsubsection{Rain and ManageRain}
Despite the name, the \textsl{Rain} class represents all the kind of precipitations (e.g. rain, hail, snow, etc.).
Similarly to the previous class, \textsl{Rain} owns a proper list of sounds, a particle system, a scene node and a reference to the camera system.

To increase the efficiency of this system, each precipitation is not executed over the whole scene, but just over the box containing the scene node the camera is currently pointing to.
This way, precipitations just follow the character and the camera, increasing performances.

\begin{figure}
	\includegraphics[width=\columnwidth]{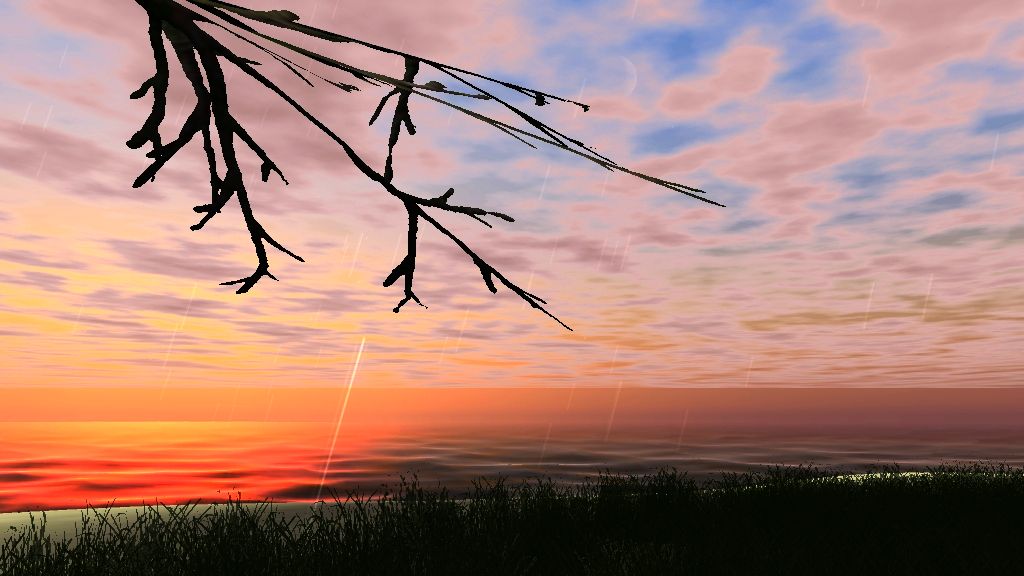}%
	\caption{The effect of raining}%
	\label{hdr7_preview}%
\end{figure}

This class exposes methods to dynamically manage atmospheric events, to configure the precipitation particle system at run-time and, finally, to reproduce sounds connected to precipitations.
Figure \ref{hdr7_preview} is a snapshot of a rainy panorama.

\subsection{Dynamic Simulation of Fluids}
\begin{figure}
	\includegraphics[width=\columnwidth]{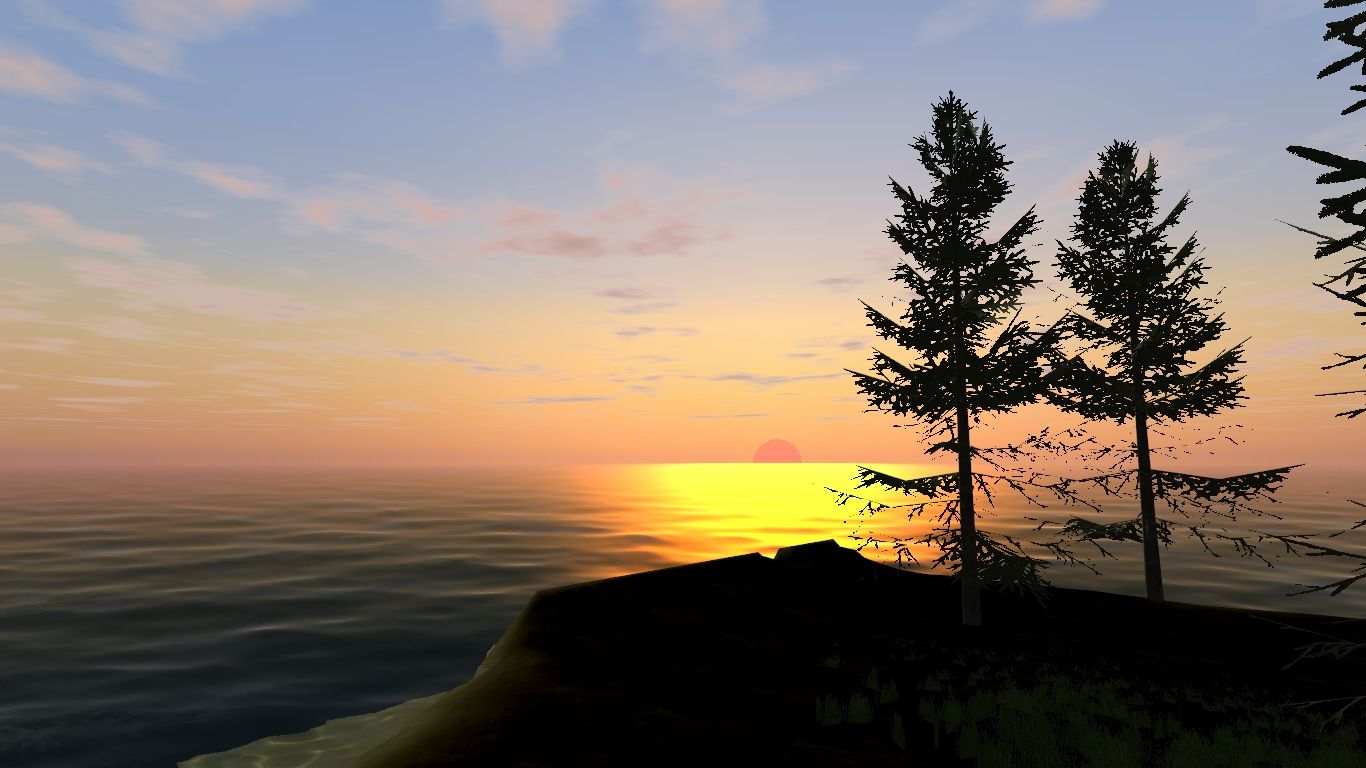}%
	\caption{Lighting and reflection on water}%
	\label{hdr12_preview}%
\end{figure}

Several work \cite{klingner2006fluid,wicke2009modular} covers the problem of simulating dynamic fluids in videogames; moreover, some commercial middleware engines have been specifically developed to manage fluid dynamics (e.g. \textsl{HydroEngine} \footnote{http://www.darkenergydigital.com/hydroengine.php}).
In our project we integrated the \textsl{Hydrax} library, an open-source solution adopted to simulate the dynamics of water and, potentially, other fluids.
In our show-case it performs at its best because the whole scene is located in a peninsula surrounded by the ocean (see Figures \ref{hdr12_preview} and \ref{hdr3_preview} for some eye-candy effects).

\subsubsection{The Hydrax Library}
\textsl{Hydrax} is developed as an open-source add-on for \textsl{OGRE}, which provides an intuitive library in order to render photo-realistic water scenes.
This library is highly configurable, thus the most of the effects which do not directly depend on shaders can be generated and managed in real-time.
This includes water depth and foam effects, smooth transitions and caustics, underwater light rays, etc.

Our improvements to the library include the possibility to manage the RTT (``Render To Texture'') texture quality (e.g. to apply additional effects before displaying the final texture), the \textsl{Hydrax} geometry (e.g. in order to calculate atmospheric effects on the water), etc.
This was possible because \textsl{Hydrax} is based on a modular interface which supports several different water geometries: i) the infinite ocean module (i.e. a projected-grid-based algorithm); ii) the simple-grid module; iii) the radial-grid module.

\subsubsection{HydraxManager}
This is the wrapper class we developed to integrate the library in the framework.
Moreover, it acts as an interface connecting \textsl{Hydrax} with the \textsl{CaelumManager} and the \textsl{WeatherManager}.
Finally, it connects the library with other fundamental modules of the framework like the scene manager, the camera, etc.

\textsl{HydraxManager} provides several methods to manage the grids (i.e. projected, simple and radial ones).

\subsubsection{Caelum and Weather on Water}
The interaction among \textsl{Hydrax} and other components is managed by two methods we introduced from scratch, \textsl{updateWeatherOnWater()} and \textsl{updateCaelumOnWater()}.

Their update is synchronized with the \textsl{timeScale} factor of the simulation, thus reducing the overload of the renderer without burdening on the graphics quality level and performances.
The first method deals with managing the effects of interaction between the dynamic skydome and the water.
For example it is involved in computing the reflection of the sun on the ocean, while it moves (thus, simulating the day-night cycle) and changes its position, color and intensity.

The latter, instead deals with simulating the tidal wave as a function of the weather conditions; thus, requiring the computing of the interaction between the wind (and, possibly, other atmospheric agents) with the water.

\subsection{Outdoor Scenes}
Industrial tools like \textsl{speedtree} \footnote{http://www.speedtree.com/} are specifically designed to deal with large-scale outdoor environments.
There are several techniques which can be adopted to manage such large virtual environments; for example, it is possible to dynamically load only the specific areas of the scene which are visible to the character.
Another interesting approach is to dynamically balance the level of details (LOD) of the objects on the scene, with respect to the distance from the camera; intuitively, it is unnecessary to reproduce high quality objects which are hardly visible because far from the camera; this way, it is possible to reduce the depth of field in order to decrease the number of objects to be rendered.
To efficiently manage outdoor scenes we exploited the \textsl{Paged Geometry} library, extending some limited features and integrating them within the framework.
These features improve the realism of reproduced outdoor environments, thus allowing developers to design large scale virtual worlds.

\subsubsection{The Paged Geometry Library}
The \textsl{Paged Geometry} engine is an open-source add-on to \textsl{OGRE} which provides highly optimized methods for rendering massive amounts of small meshes, covering a possibly infinite area. 
It is particularly suited for representing forests and outdoor scenes, containing millions trees, grass, rocks, etc.
\textsl{Paged Geometry} introduces the main advantage, with respect to plain entities, of allowing a dynamic balancing of the level of detail.
This efficient approach ensures better performances in particular with outdoor scenes, which can benefit of a frame rate increase of an order of magnitude.
The algorithm relies on paging the geometry: only entities which are immediately visible are loaded. 
This drastically reduces the memory requirement and avoids memory leaks for very large outdoor scenes.
In details, \textsl{Paged Geometry} relies on three different \emph{paging systems}: i) \textsl{BatchPage}; ii) \textsl{WindPage}; iii) \textsl{ImpostorPage}. 
The first is adopted to render elements near to the camera; it supports the dynamic lighting system.
The second is similar and supports the animation of effects of the wind on the foliage.
The latter is adopted to render objects far from the camera, using \emph{static billboards} instead of three-dimensional meshes; this system does not natively support the dynamic lighting, thus introducing graphics artifacts in an environment completely based on this lighting system.
We extended the library to support the dynamic lighting of \textsl{ImpostorPage}-based elements.

\subsubsection{PagedGeometryManager}
We developed a wrapper class, namely \textsl{PagedGeometryManager}, which integrates the library within the framework.
This class provides connection to the scene manager, the camera manager, and to objects to be rendered.
It deals in particular with the initialization of two components, the \textsl{TreeLoader} and the \textsl{GrassLoader}.
Finally, we defined methods to provide support to dynamic lighting, a feature not natively supported by the engine.

\subsubsection{TreeLoader and GrassLoader}
\begin{figure}
	\includegraphics[width=\columnwidth]{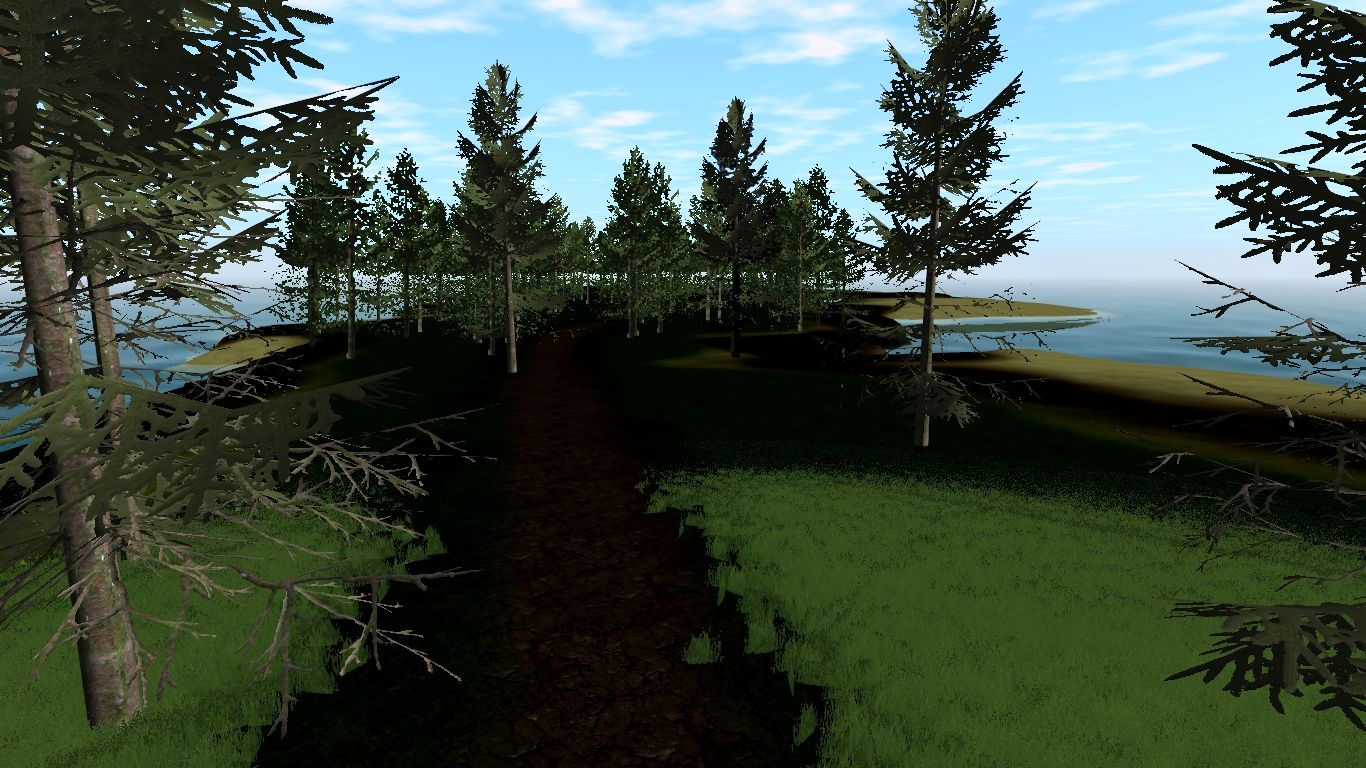}%
	\caption{LOD balancing and paged geometry}%
	\label{hdr11_preview}%
\end{figure}

The first technique introduced in the library is the random generation of a 'forest', adopting density maps (Figure \ref{maps}, top right).
A density map is a grayscale image whose areas represent positions of the map in which the \textsl{TreeLoader} and the \textsl{GrassLoader} can dynamically instantiate objects.
Our \textsl{TreeLoader} supports the random generation of a plethora of defined tree meshes, in a random scale lying in a given interval.
The original library provides methods for setting an infinite grid or, otherwise, some boundaries for paging the size of blocks which represent the grid map. 
The level of detail is configured via methods to manage the \textsl{BatchPage}, \textsl{WindPage} and \textsl{ImpostorPage} sizes.
The \textsl{GrassLoader}, instead, adopts only one level of detail (\textsl{GrassPage}).
In Figures \ref{hdr11_preview} and \ref{hdr3_preview}, we show the dynamic level of detail balancing.

\subsubsection{Dynamic Lighting}
\begin{figure}
	\includegraphics[width=\columnwidth]{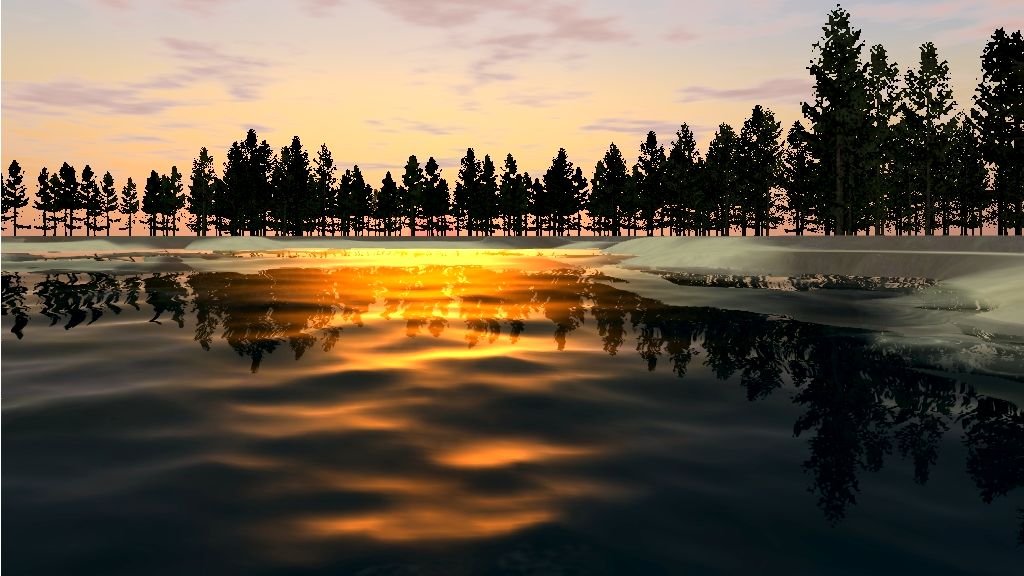}%
	\caption{Dynamic lighting and fluid dynamics}%
	\label{hdr3_preview}%
\end{figure}

We introduced the support to dynamic lighting in \textsl{Paged Geometry}, in particular on entities based on the \textsl{ImpostorPage} paging system.
This technique is implemented using shaders. 
Our algorithm relies on the \emph{Per-Vertex} lighting model, implemented on four light components: i) emissive light; ii) ambient light; iii) diffuse light and, finally, iv) specular light. 
Dynamic lighting affects reflections (Figure \ref{hdr3_preview}).

\section{Conclusions}
\label{conclusion}
In this work we introduced novel extensions to our framework for supporting the game development process.
It completely relies on state-of-the-art open-source solutions.
All the source-code has been released as open-source for possible further improvements \footnote{http://informatica.unime.it/velab/}.
Our first contribution consists of improving existing libraries.
We provided an integration interface to several components, obtaining, de facto, a valid middleware framework for supporting the development of 3D videogames set in virtual environments.
Moreover, we developed from scratch a library providing features for the simulation of weather conditions in this environment.
This work is testified by developing and illustrating a show-case demonstration which depicts some of the features our framework provides.
This project could be still extended introducing new features.
Some future work will focus on increasing even more the degree of realism of the rendering engine, including new rendering techniques such as the \emph{tessellation} \cite{li2010method}.
Additional environmental and atmospheric agents will be included, e.g. effects of the wind such as tornadoes, sandstorms, etc.
Another important aspect to be introduced is the volumetric cloud simulation \cite{stiver2010sketch}, thus improving the realism of these atmospheric agents.


\section*{Acknowledgments}
Valerio Mazzeo contributed within his graduation dissertation to the design and implementation of this application.

%
\bibliographystyle{abbrv}
\bibliography{sigproc-disio2}  
%
%

\end{document}